\documentclass[acmtog]{acmart}

\AtBeginDocument{%
  }

\usepackage{subfigure}
\usepackage{bm}
\usepackage{amsmath}

\usepackage{color}

\copyrightyear{2025}
\acmYear{2025}
\setcopyright{acmlicensed}\acmConference[SA Conference Papers '25]{SIGGRAPH Asia 2025 Conference Papers}{December 15--18, 2025}{Hong Kong, Hong Kong}
\acmBooktitle{SIGGRAPH Asia 2025 Conference Papers (SA Conference Papers '25), December 15--18, 2025, Hong Kong, Hong Kong}
\acmDOI{10.1145/3757377.3763894}
\acmISBN{979-8-4007-2137-3/2025/12}

\acmSubmissionID{papers\_1586}

\newcommand{\blue}{\textcolor{blue}}

\newcommand{\del}[1]{}
\renewcommand{\color}[1]{}

\begin{document}

\title {HRM\textsuperscript{2}Avatar: High-Fidelity Real-Time Mobile Avatars from Monocular Phone Scans}

\author{Chao Shi}
\authornote{Both authors contributed equally to this research.}
\affiliation{%
  \institution{Alibaba Group}
  \city{Hangzhou}
  \country{China}
}

\author{Shenghao Jia}
\authornotemark[1]
\affiliation{\institution{Shanghai Jiao Tong University}
\city{Shanghai}
\country{China}}
\affiliation{\institution{Alibaba Group}
\city{Hangzhou}
\country{China}}

\author{Jinhui Liu}
\affiliation{%
  \institution{Alibaba Group}
  \city{Hangzhou}
  \country{China}
}

\author{Yong Zhang}
\authornote{Corresponding Author.}
\affiliation{%
  \institution{Alibaba Group}
  \city{Hangzhou}
  \country{China}
}
\email{guyu.zy@taobao.com}

\author{Liangchao Zhu}
\affiliation{%
  \institution{Alibaba Group}
  \city{Hangzhou}
  \country{China}
}

\author{Zhonglei Yang}
\authornote{Project Leader.}
\affiliation{%
  \institution{Alibaba Group}
  \city{Hangzhou}
  \country{China}
}

\author{Jinze Ma}
\affiliation{%
  \institution{Alibaba Group}
  \city{Hangzhou}
  \country{China}
}

\author{Chaoyue Niu}
\affiliation{%
  \institution{Shanghai Jiao Tong University}
  \city{Shanghai}
  \country{China}
}

\author{Chengfei Lv}
\authornotemark[2]
\affiliation{\institution{Alibaba Group}
\city{Hangzhou}
\country{China}}
\email{chengfei.lcf@taobao.com}

\renewcommand{\shortauthors}{Shi et al.}

\begin{abstract}
We present HRM$^2$Avatar, a novel framework for creating high-fidelity avatars from monocular phone scans, which can be rendered and animated in real-time on mobile devices.
Monocular capture with commodity smartphones provides a low-cost, pervasive alternative to studio-grade multi-camera rigs, making avatar digitization accessible to non-expert users.
Reconstructing high-fidelity avatars from single-view video sequences poses significant challenges due to deficient visual and geometric data relative to multi-camera setups.
To address these limitations, at the data level, our method leverages two types of data captured with smartphones: static pose sequences for detailed texture reconstruction and dynamic motion sequences for learning pose-dependent deformations and lighting changes.
At the representation level, we employ a lightweight yet expressive representation to reconstruct high-fidelity digital humans from sparse monocular data. 
First, we extract explicit garment meshes from monocular data to model clothing deformations more effectively. 
Second, we attach illumination-aware Gaussians to the mesh surface, enabling high-fidelity rendering and capturing pose-dependent lighting changes.
This representation efficiently learns high-resolution and dynamic information from our tailored monocular data, enabling the creation of detailed \del{and lifelike} avatars.
At the rendering level, real-time performance is critical for rendering and animating high-fidelity avatars in AR/VR, social gaming, and on-device creation, demanding sub-frame responsiveness. Our fully GPU-driven rendering pipeline delivers 120 FPS on mobile devices and 90 FPS on standalone VR devices at 2K resolution, over $2.7\times$ faster than representative mobile-engine baselines.
Experiments show that HRM$^2$Avatar delivers superior visual realism and real-time interactivity at high resolutions, outperforming state-of-the-art monocular methods.
\end{abstract}



\begin{teaserfigure}
  \includegraphics[width=\textwidth]{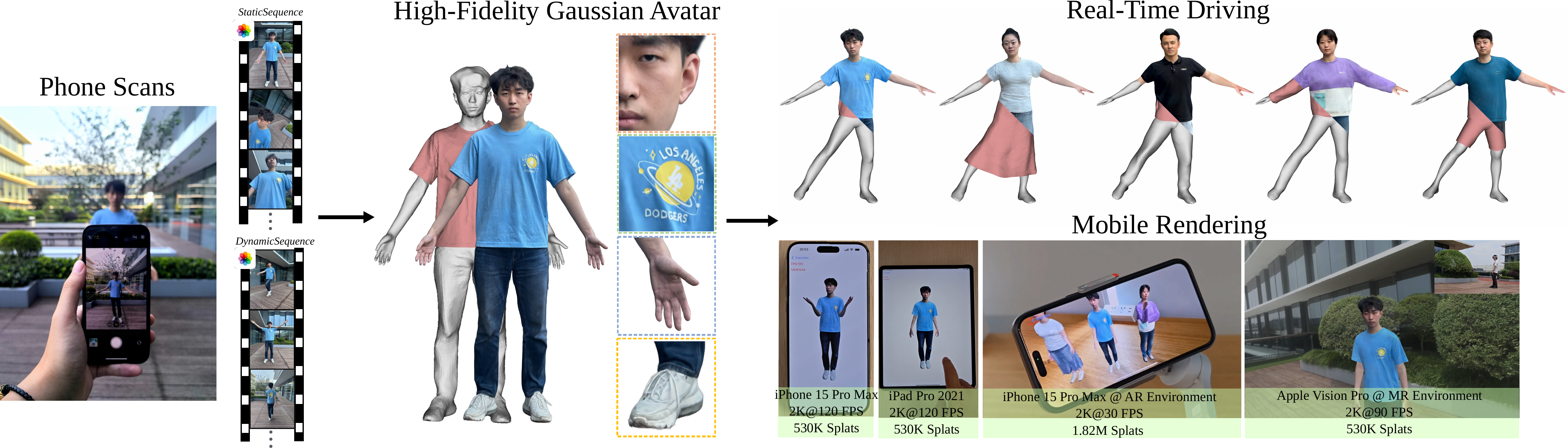}
  \caption{Our method creates high-fidelity avatars with realistic clothing dynamics by monocular smartphone scanning,
  and achieves 2048×945@120FPS on iPhone 15 Pro Max and 1920×1824x2@90FPS on Apple Vision Pro with 533,695 splats. 
  Each subject's data is captured using a single iPhone for 5 minutes.
  }
  \Description{Please see the caption.}
  \label{fig:teaser}
\end{teaserfigure}

\maketitle

\section{INTRODUCTION}
High-fidelity reconstruction, animation and real-time rendering of full-body human avatars are pivotal for interactive applications including online meetings, filmmaking, gaming, augmented reality (AR) and virtual reality (VR). 
Enabling users to generate high-fidelity avatars from accessible monocular smartphone scans and drive them on mobile devices has practical impact for immersive social and collaborative experiences. %
Existing methods based on Neural Radiance Fields (NeRF)~\cite{Nerf} and 3D Gaussian Splatting (3DGS)~\cite{kerbl20233Dgaussians}, leveraging parameterized body priors such as SMPL-X~\cite{SMPX2019Pavlakos}, have achieved full-body human avatar reconstruction from monocular inputs~\cite{Jiang2022SelfRecon, yu2023monohuman, hu2024gaussianavatar, guo2025vid2avatarpro, moon2024exavatar}. 
However, these approaches struggle to maintain high-resolution fidelity, capture fine-grained motion details and enable real-time driving on mobile devices.
Specifically, reconstructing high-fidelity animatable avatars from monocular inputs for mobile applications is constrained by three critical challenges: 
\begin{itemize}
\item \textbf{Limited visual detail in monocular reconstructions.} Fine-grained details such as intricate fabric textures and skin microstructures are lost in the captured images due to the necessity of capturing the full-body at significant distances for robust pose estimation. 
\item \textbf{Inadequate modeling of dynamic deformations and illumination variations.} Dynamic deformations, encompassing body and clothing deformations and their relative interactions, are often modeled monolithically, resulting in blurred garment boundaries or distorted kinematics.
\item \textbf{Computational bottlenecks in high-resolution rendering pipelines.} 
Despite the demand for real-time immersive experiences, achieving interactive frame rates on mobile hardware remains challenging due to the computational intensity of photorealistic rendering with NeRF or 3DGS.
\end{itemize}

To address these challenges, we present \textbf{HRM$^2$Avatar}, an end-to-end framework that generates high-fidelity clothed full-body avatars, explicitly modeling non-rigid deformation and illumination variations from monocular smartphone captures, and enabling real-time and high-resolution interaction on mobile devices including AR/VR headsets, as presented in Fig.~\ref{fig:teaser}. 
The framework begins with an accessible monocular image sequences scanning process, capturing both static and dynamic information of the subject.
The \del{clothed} avatar is represented by a \del{multi-layer} {\color{blue} clothed} mesh-driven Gaussian Splatting framework.
Non-rigid deformations and pose dependent illumination variations are explicitly modeled and distilled to lightweight Multi-Layer Perceptrons (MLPs), 
ensuring high-fidelity reconstruction and realistic animation. 
A static-dynamic co-optimization strategy jointly refines texture details from static close-ups and dynamic deformations and illumination variations from motion sequences. This strategy mitigates overfitting risks in sparse monocular data while preserving fine-grained realism.
For real-time deployment on mobile devices, the mesh-driven Gaussian rendering pipeline is specifically optimized, achieving speedup of $4.01\times$ compared to the Unity implementation~\cite{UnityGS2023Aras} and $2.74\times$ compared to the Godot implementation~\cite{godotGS2023}.
Our contributions are as follows:
\begin{itemize}
    \item We introduce an end-to-end mobile avatar creation and driving system, which takes %
    monocular smartphone captures as input, reconstructs both full-body appearance and close-up details with high fidelity, and supports real-time driving on mobile devices.
    \item We introduce a \blue{clothed} mesh-Gaussian hybrid framework for avatar representation, integrating pose-dependent geometric deformation and illumination variation to enable dynamic and high-fidelity avatar reconstruction.
    \item A customized GPU-driven rendering pipeline integrating data rearrangement, 
    hierarchical culling and single-pass stereo rendering is developed for mobile devices, achieving high-resolution real-time performance (e.g., 2048×945 @ 120 FPS on iPhone 15 Pro Max and 1920×1824x2 @ 90 FPS on Apple Vision Pro with 533,695 splats). {\color{blue} The code and sample assets are available at \url{https://acennr-engine.github.io/HRM2Avatar}}.
\end{itemize}

\begin{figure*}[!t]
\centering
\includegraphics[width=0.95\linewidth]{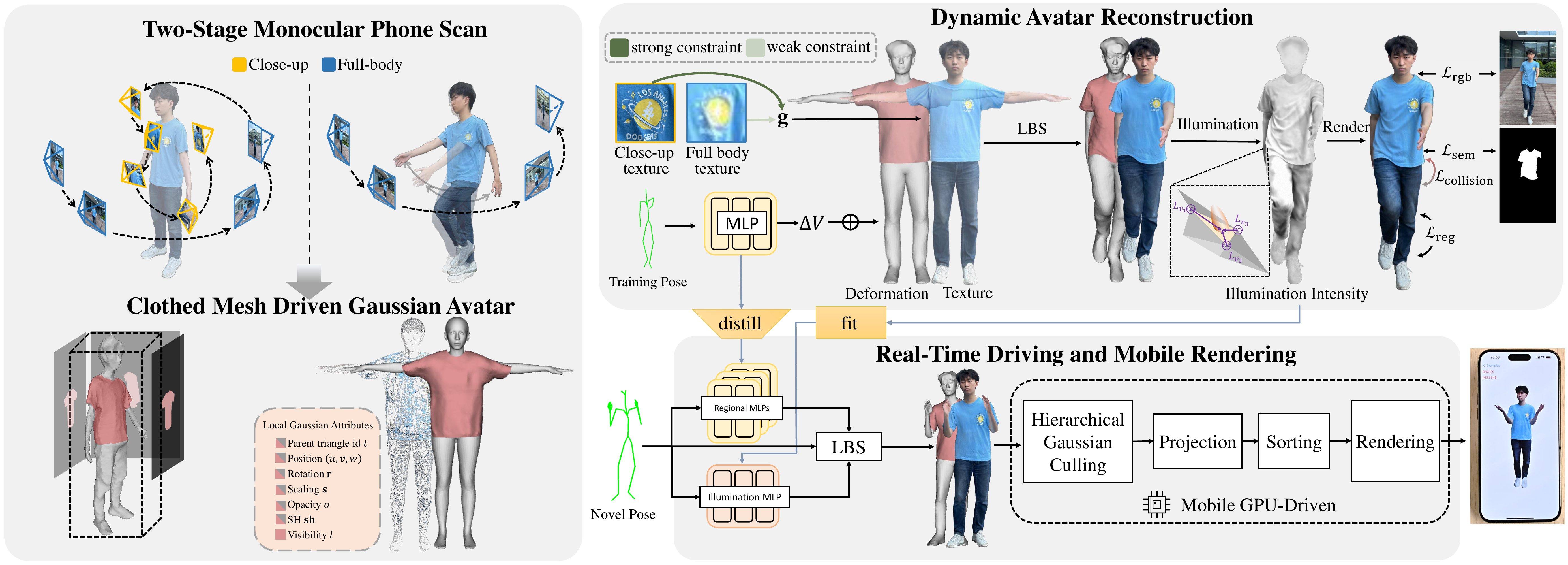}
\caption{\textbf{Method Overview.} Given the two-stage phone scans of a subject, we construct a clothed mesh-driven Gaussian avatar. Static, texture-rich images impose stringent supervision on Gaussian attributes $\mathbf{g}$, while dynamic, motion-intensive sequences prioritize optimization of deformation $\Delta \mathbf{V}^d$ and illumination $L$. Through deformation MLP, illumination MLP and GPU-driven Gaussian rendering pipeline, real-time rendering and animation of realistic avatars is achievable on mobile devices.}
\label{overview}

\Description{Please see the caption.}
\end{figure*}

\section{RELATED WORK}
\paragraph{Monocular Full-Body Avatar Reconstruction.} 
Methods based on traditional mesh-texture rendering can reconstruct human body meshes from monocular video inputs~\cite{SMPX2019Pavlakos, DeepCap2020Habermann}, but struggle to reproduce photorealistic avatars due to insufficient texture detail and limited dynamic expressiveness. 
NeRF-based methods~\cite{guo2023vid2avatar, Jiang2024MultiPly, weng2022humannerf} map query points into a canonical space using inverse skinning to reconstruct high-fidelity avatars from monocular videos, but their implicit representations limit pose controllability and real-time rendering. 
3DGS-based methods address these limitations via binding splats to human body meshes (e.g., SMPL/SMPL-X) and optimizing 3D Gaussian attributes~\cite{Lei2024GART, SplattingAvatar2024Shao} or regressing Gaussian parameters via neural networks~\cite{kocabas2024hugs, moon2024exavatar, hu2024gaussianavatar, Qian20243DGS-Avatar}.
Recent approaches adopt joint optimization of surface meshes and 3D Gaussian splats, leveraging the predefined topology of meshes to improve deformability~\cite{moon2024exavatar, qian2024gaussianavatars}. 
Despite these advancements, existing methods are limited by their reliance on full-body input videos, resulting in degraded quality for close-up details.
Alternative approaches employing generative models to infer unseen images~\cite{ho2024sith, Xiang2024OneShot}, or prior model to regress Gaussian attributes~\cite{guo2025vid2avatarpro, qiu2025LHM} struggle to maintain global 3D consistency and require high-quality training data to create high resolution avatars. 

\paragraph{Clothed Avatar Reconstruction.} 
Most existing full-body avatar methods model clothing and the body as a monolithic entity~\cite{Burov2021Dynamic, Li2025PoseVocab,guo2023vid2avatar, hu2024gaussianavatar, Qian20243DGS-Avatar, guo2025vid2avatarpro, Li2024AnimatableGaussians}. While this simplifies implementation, it cannot capture fine-grained clothing dynamics.
To better represent clothing properties on top of the bodies, especially the relative motion, recent approaches model clothing as disentangled layers~\cite{Xiang2022DressingAvatars, Modeling2021Xiang, Lin2024LayGA, Chen2025TaoAvatar, zielonka25dega}, using multi-view capture systems to reconstruct physically plausible interactions.
However, these methods require costly multi-view systems, limiting widespread adoption.
In monocular setups, imposing constraints on clothing is highly challenging. SCARF~\cite{Feng2022Capturing} and DELTA~\cite{Feng2023DELTA} extract individual implicit NeRF-based clothing, while GGAvatar~\cite{Chen2024GGAvatar} employs the Implicit Surface Prediction (ISP) Model~\cite{Li2023isp} to extract separate clothing mesh and attach Gaussians for photorealistic reconstruction. 
Although these methods circumvent the drawbacks of monolithic representation, their reconstructions
do not meet high-resolution requirements due to insufficient detail preservation and dynamic deformation modeling.

\paragraph{Efficient Gaussian Avatar Rendering on Mobile Devices.}
NeRF-based approaches incur high computational costs due to ray-marching requirements, hindering real-time performance. {\color{blue} Methods like MobileNeRF~\cite{chen2022mobilenerf} and Binary Opacity Grids~\cite{Reiser2024BinaryOpacityGrids} improve rendering efficiency on mobile devices, but they remain limited to static scene rendering and have only demonstrated real-time performance at relatively low resolutions.}
While 3DGS enables real-time rendering of static objects on desktops, animating avatars remains computationally intensive~\cite{Qian20243DGS-Avatar, Pang2024ASH,kocabas2024hugs, zhan2025realtimehighfidelitygaussianhuman}.
Recent optimizations include 
LoDAvatar's~\cite{dongye2024lodavatar} hierarchical embedding and adaptive levels of detail (LOD) (262 FPS on PC) and FlashAvatar's~\cite{xiang2024flashavatar} lightweight representations achieving 300 FPS at $512\times512$ resolution. 
Despite these advances, mobile deployment remains challenging due to hardware constraints: for instance,  SplattingAvatar~\cite{SplattingAvatar2024Shao} drops from 300 FPS on PC to 30 FPS on iPhone 13;
TaoAvatar~\cite{Chen2025TaoAvatar} achieves 4D avatar rendering %
with 90 FPS on Apple Vision Pro and 60 FPS on the Android device, but restricts avatars to precomputed poses.
SqueezeMe~\cite{saito2024squeezeme} enables concurrent rendering  of three avatars at 72 FPS on Meta Quest 3 using UV-space Gaussian location, linear distillation and Gaussian corrective sharing, though 
visual fidelity degrades in articulated regions (e.g., arms, hands) due to the 60,000-splat-per-avatar limit.
Our work further enhances Gaussian avatar rendering efficiency on mobile platforms via an optimized GPU-driven rendering pipeline, enabling real-time interactive driving of high-fidelity avatars with 530K splats at 120 FPS.

\section{METHOD}
\looseness-1We present \textbf{HRM\textsuperscript{2}Avatar}, an end-to-end system that reconstructs a high-fidelity, fully animatable human avatar from a single-phone scan and achieves real-time rendering on mobile GPUs at 2K resolution, as shown in Fig.~\ref{overview}. 
\textbf{Stage 1 — Capture (Sec.~\ref{subsec:data capture}):} \textit{StaticSequence} records intricate visual details, while \textit{DynamicSequence} captures movement for pose-dependent deformation and illumination learning;
\textbf{Stage 2 — Representation (Sec.~\ref{section: AVATAR MESH REPRESENTATION}):} {\color{blue} a clothed mesh-driven} Gaussian model pairs explicit garment meshes for cloth dynamics with illumination-aware 3D Gaussians for appearance; 
\textbf{Stage 3 — Optimization (Sec.~\ref{section: 4D Dynamic Avatar Reconstruction}):} geometry, texture, and per-frame lighting are alternately refined across both sequences, fitting illumination maps to decouple lighting from shape; 
\textbf{Stage 4 — Rendering (Sec.~\ref{section: AVATAR DRIVING}):} a GPU-Driven Gaussian avatar rendering pipeline with mesh-guided hierarchical culling, in-place data rearrangement, and single-pass stereo output sustains real-time performance on mobile hardware.

\subsection{Two-stage Monocular Data Capture}\label{subsec:data capture}
Monocular video inherently lacks explicit depth cues and view-dependent appearance variations, posing significant challenges for 3D reconstruction and photorealistic rendering.
To overcome these limitations, we propose a two-stage smartphone-based scanning protocol that captures complementary geometric, textural, and dynamic information.  
We introduce a dual-sequence capture process for each subject, comprising \textit{StaticSequence} and \textit{DynamicSequence}: 
\begin{itemize}
\item \textit{StaticSequence}: the subject maintains a stationary A-pose, a stable and easy-to-maintain configuration. The camera operator first orbits around the subject to capture full-body images $I_{sg}$. Then the photographer captures close-up images $I_{sl}$ of texturally rich localized details, such as apparel logos and hands, without requiring the entire body visible within the frame.%
Notably, our method accommodates minor subject movements, ensuring robust reconstruction. 

\item \looseness-1\textit{DynamicSequence}: the subject performs articulated motions, including arm elevation, elbow flexion, and leg elevation, designed to capture non-rigid deformations and pose-dependent shadows across primary joint rotations. 
The camera operator orbits the subject to acquire full-body images $I_{d}$ during these motions. 

\end{itemize}

The \textit{StaticSequence} provides detailed references for clothing and skin textures and facilitates the extraction of garment meshes for subsequent reconstruction and animation.
The \textit{DynamicSequence} captures dynamic information, including non-rigid deformations and variations in lighting and shadows. Together, these sequences yield approximately 300-400 images per subject, balancing reconstruction accuracy and capture efficiency. {\color{blue} The capturing strategy is applicable to most clothing types. For clothes with more complex shapes, it may be necessary to add extra poses for \textit{DynamicSequence} to reveal parts that are occluded in regular poses.}

\subsection{Clothed Mesh-Driven Gaussian Avatar Representation} \label{section: AVATAR MESH REPRESENTATION}
To enable high-fidelity animation of clothed avatars in monocular reconstruction scenarios, we propose a hybrid mesh-Gaussian representation that decouples body and clothing dynamics while ensuring geometric consistency. We extend the SMPL-X body model with explicit garment meshes, forming a clothed SMPL-X representation, and bind Gaussians to the mesh triangles for photorealistic rendering across arbitrary motions. 

\paragraph{Preprocess and Clothed Body Registration}
We assume the subject remains stationary during the \textit{StaticSequence} to derive initial camera and SMPL-X pose parameters.
We employ COLMAP~\cite{schoenberger2016sfm, schoenberger2016mvs} to estimate camera parameters for all \textit{StaticSequence} images, particularly the relative camera positions between full-body ($I_{sg}$) and close-up ($I_{sl}$) images.
Under this assumption, full-body images $I_{sg}$ share the same SMPL-X parameters, which is estimated by an off-the-shelf SMPL-X regressor (for the first frame)~\cite{shen2024gvhmr, pavlakos2024reconstructingHands, Moon2022Hand4Whole} and finetuned with detected 2D keypoints(for all full-body images).
Due to challenges in robust SMPL-X parameter estimation for close-up images, we allow close-up images to inherit the globally optimized body parameters, and register them to corresponding body regions using estimated camera extrinsic parameters.
To mitigate biases from minor subject movements, we apply frame-wise residual corrections to camera and pose during training, optimizing these corrections to ensure geometric consistency across frames. For \textit{DynamicSequence}, we estimate SMPL-X parameters for each frame independently.

To initialize clothing, we employ NeuS2~\cite{neus22023Wang} to reconstruct the clothed body geometry from \textit{StaticSequence} images, and segment clothing components via semantic-supervised differentiable rendering ~\cite{khirodkar2024sapiens, Laine2020diffrast}. 
To animate the extracted clothing mesh, we transfer skinning weights from the estimated SMPL-X model to the clothing mesh via nearest-point matching~\cite{Bertiche2021PBNS}. 
Using inverse linear blend skinning (LBS), clothing is transformed back to align with SMPL-X’s T-pose. The integration of body and clothing produces the clothed SMPL-X model, a comprehensive personalized parametric representation.

\paragraph{Drivable Gaussian Binding.} 
Gaussians are bound to mesh triangles of clothed SMPL-X model to encode photorealistic appearance while enforcing geometric constraints.
Each Gaussian is parameterized by local attributes \(\textbf{g}=\{t, (u, v, w), \textbf{r}, \textbf{s}, o, \textbf{sh}, l\}\), where \(t\) denotes the index of the parent triangle.
The parameters \((u, v, w)\), \(\textbf{r}\), and \(\textbf{s}\) denote the center position, rotation, and scaling within the parent triangle's local space.
Specifically, \(u\) and \(v\) represent barycentric coordinates, and \(w\) indicates the offset distance of the Gaussian center along the triangle normal.
The attribute \(\textbf{o}\) represents opacity, \(\textbf{sh}\) denotes the spherical harmonic (SH) coefficients, and \(l\) is the discrete visibility label, indicating single-face visible Gaussian which is detailed in supplementary material.
To manage Gaussian density, we split oversized Gaussians and clone undersized ones, following ~\cite{kerbl20233Dgaussians}.
Newly generated Gaussians inherit the $t$ and $l$ attributes 
from their parent Gaussians. 
We adopt the SurFhead method ~\cite{lee2024surfheadaffinerigblending} for local-to-global Gaussian transformations, enabling stretching and shearing to adapt to changes in triangle geometry.
In non-hair regions, we constrain Gaussians to two dimensions on the mesh surface by setting their normal-direction scale and offset to zero, permitting only rotation about the normal.
{
\color{blue}
Such configuration prevents penetration through clothing layers, reduces in-motion artifacts such as spikes, and preserves reconstruction clarity. 
It should be noted that we compensate for geometry inaccuracies by refining the reconstructed mesh using gradients from Gaussian Splatting differentiable rendering. Instead of applying per-Gaussian offset adjustments, this approach achieves accurate and plausible rendering without sacrificing visual consistency.
}

{\color{blue}
We choose to bind 2D Gaussian splats to the clothed mesh rather than employ texture-based mesh representations. Texture maps generated through differentiable rendering are prone to UV seam discontinuities arising from mipmap-based sampling. Additionally, single-layer textured meshes tend to produce unnaturally thin, paper-like garment edges at critical features such as cuffs, collars, and hemlines. In contrast, Gaussian splats inherently blend beyond mesh boundaries, enabling the capture of realistic silhouettes.
}

\subsection{Dynamic Avatar Reconstruction} 
\label{section: 4D Dynamic Avatar Reconstruction}
To mitigate ambiguities in monocular data and achieve high-fidelity animatable avatars,  we propose a static-dynamic co-optimization framework that decouples illumination modeling, texture optimization, and deformation learning.
This framework explicitly models pose-dependent deformations and illumination variations, leveraging complementary constraints from static and dynamic sequences.  

\paragraph{Shape Reconstruction.} 
The clothed SMPL-X model generates a posed 3D human-clothing mesh $\mathbf{V}$ via LBS:
\begin{equation}
\mathbf{V} = \text{LBS}\left( 
\mathbf{V}_{\text{T}}
 + \Delta \mathbf{V}, \mathbf{\boldsymbol{\theta}} \right),
\end{equation}
where $\mathbf{V}_{\text{T}}$ represents the template vertices of the clothed SMPL-X model (comprising the shaped body and reconstructed clothing meshes), $\boldsymbol{\theta}$ denotes the estimated pose parameters, and $\Delta \mathbf{V}$ captures non-rigid deformations, such as cloth wrinkles and soft tissue movements, beyond skeletal deformation

We optimize $\Delta \mathbf{V}$ through inverse rendering using 3D Gaussians. Specifically, deformation offsets are computed in the LargeSteps~\cite{nicolet2021large} optimization framework to enhance convergence and robustness, then mapped to Euclidean space for final shape reconstruction, and $\Delta \mathbf{V}$ is divided into three parts as
\begin{equation}
    \Delta \mathbf{V} = LS(\Delta \mathbf{V}^{s}) + LS(\Delta \mathbf{V}^{d}(\boldsymbol{\theta})) + \Delta \mathbf{V}^{f},
\end{equation}
where \(\Delta \mathbf{V}^{s}\) captures pose-independent offsets such as hairstyles, footwear, and clothing misalignment, while \(\Delta \mathbf{V}^{d}(\boldsymbol{\theta})\) represents pose-dependent offsets regressed via a multi-layer perceptron (MLP) for each vertex based on body pose \(\boldsymbol{\theta}\) and vertex coordinate in canonical space.
Both \(\Delta \mathbf{V}^{s}\) and \(\Delta \mathbf{V}^{d}(\boldsymbol{\theta})\) are formulated within the LargeSteps space and $LS(\cdot)$ denotes the transformation to Euclidean space.
Due to clothing dynamics, such as swinging or flapping motions, which induce deformations influenced by both current pose and motion history, we introduce a frame-wise compensation term $\Delta \mathbf{V}^{f}$ to model these pose-independent deformations explicitly.  
During training, \(\Delta \mathbf{V}^{d}(\boldsymbol{\theta})\) rapidly converges to capture most pose-dependent offsets, while smaller perturbations are addressed by the frame-wise compensation \(\Delta \mathbf{V}^{f}\).

\paragraph{Illumination Modeling.} %
Prior studies~\cite{moon2024exavatar, qian2024gaussianavatars} model pose-conditioned Gaussian colors directly using neural networks to capture illumination variations.
However, these neural networks are prone to overfitting due to the inherent data sparsity in monocular captures, compromising generalization performance.
Instead, we explicitly model a single-channel illumination intensity conditioned on pose. 
Empirical analysis shows that pose-related illumination changes, driven by alterations in surface normal orientation and shadows from occlusions and wrinkles, primarily manifest as brightness modulations rather than chromatic variations.  

Specifically, the color of Gaussian \(i\) for frame \(f\) is expressed as
\begin{equation}
    \mathbf{c}_i^{f}(\mathbf{d}) = \phi(\mathbf{sh}_i, \mathbf{d}) \cdot L_i^{f},
\end{equation}
where \(\phi(\mathbf{sh}_i, \mathbf{d})\) samples the spherical harmonic (SH) coefficients \(\mathbf{sh}_i\) in direction \(\mathbf{d}\),
and \(L_i^{f}\) is the illumination intensity.
This formulation draws from modern rendering techniques, such as ambient occlusion (AO) and shadows, which modulate brightness due to occlusion. 
Considering the spatial continuity of illumination variations, we interpolate the illumination intensity for Gaussian \(i\) from the intensities at the three vertices of its corresponding triangle using barycentric coordinates, fitting per-frame vertex intensities during reconstruction. Figure~\ref{overview}, top-right, illustrates an example of illumination intensity fitted to a single frame.

Our reconstruction process outputs vertex positions and pose-dependent illumination intensities for each frame.  
We train lightweight neural networks to predict vertex position offsets and illumination intensities from pose \(\theta\), minimizing the $L_1$ loss with respect to the reconstructed frames for real-time animation. {\color{blue} Note that directly training the lightweight neural networks during the reconstruction phase results in poor convergence, primarily due to their simplified architectures designed for efficient inference and the single-sample training batches inherited from the original 3DGS framework.}

\paragraph{Gradient Control.}
We jointly optimize Gaussian attributes, deformations and illumination parameters using both \textit{StaticSequence} and \textit{DynamicSequence}.
\blue{
The two sequences are inherently heterogeneous: \textit{StaticSequence} provides pose-independent Gaussian attributes, while \textit{DynamicSequence} encodes pose-dependent illumination and deformation. To balance their contributions during optimization, we introduce a dual-channel gradient propagation strategy that limits the impact of dynamic data on pose-independent Gaussian attributes.
}
Specifically, during backpropagation, we assign distinct weights to the gradients of Gaussian attributes $\textbf{g}$ for close-up images {\color{blue} $I_{sl}$} and full-body images {\color{blue} $I_{sg} \cup I_d$}:
\begin{equation}
\textbf{g}^{(t+1)} = \textbf{g}^{(t)} - \alpha \cdot \frac{\partial \mathcal{L}(I, I_{gt})}{\partial \textbf{g}}, \alpha = \begin{cases} \alpha_\text{major}, & I_{gt} \in I_{sl} \\ \alpha_\text{minor}, & I_{gt} \in I_{sg} \cup I_{d} \end{cases}.
\end{equation}
We set $\alpha_\text{major}$ greater than $\alpha_\text{minor}$ because close-up images provide more accurate static texture gradients. 
Specifically, $\alpha_\text{major}$ is set to 5 and $\alpha_\text{minor}$ to 1 consistently in all our experiments.
Additionally, we perform Gaussian splitting based solely on gradients accumulated from StaticSequence images, preventing errors from high-frequency components in dynamic data.

\paragraph{Losses.} 
The loss function integrates constraints in photometric and geometric spaces, with the former optimizing Gaussian attributes and deformations, and the latter constraining the geometric relationships between clothing and body.  
It comprises four components:  
\begin{equation}
    \mathcal{L} = \mathcal{L}_\text{rgb} + \mathcal{L}_\text{reg} + \lambda_\text{sem}\mathcal{L}_\text{sem} + \lambda_\text{collision}\mathcal{L}_\text{collision},
\end{equation}
where $\lambda_{*}$ represents loss weights, $\mathcal{L}_\text{rgb}, \mathcal{L}_\text{reg}, \mathcal{L}_\text{sem}$, and $\mathcal{L}_\text{collision}$ are detailed below.

The $\mathcal{L}_\text{rgb}$ term incorporates pixel-wise ($L1$, SSIM, mask) and perceptual-based (LPIPS~\cite{zhang2018perceptual}) photometric losses, defined as
\begin{equation}
    \mathcal{L}_\text{rgb} = \lambda_{L1}\mathcal{L}_{L1} + \lambda_\text{ssim}\mathcal{L}_\text{ssim} + \lambda_\text{lpips}\mathcal{L}_\text{lpips} + \lambda_\text{mask}\mathcal{L}_\text{mask}.
\end{equation}

The $\mathcal{L}_\text{reg}$ term regularizes Gaussian attributes and mesh geometry, with details provided in the supplementary material.  

{\color{blue} The semantic loss term $\mathcal{L}_\text{sem}$ is introduced to regularize and guide cloth geometry reconstruction.
We assign non-learnable semantic labels to each Gaussian to indicate whether it lies on the clothing mesh. $\mathcal{L}_\text{sem}$ is defined as the $L_1$ loss between the rendered mask of clothing Gaussians and the 2D clothing segmentation mask predicted by Sapiens~\cite{khirodkar2024sapiens}.
}

The $\mathcal{L}_\text{collision}$ term penalizes clothing-body collisions.
To address initialization misalignment in monocular scenes, we introduce a normal consistency constraint derived from PBNS's collision loss framework~\cite{Bertiche2021PBNS}, with detailed formulations provided in the supplementary materials.

\subsection{Real-Time Gaussian Avatar Rendering} \label{section: AVATAR DRIVING}
\begin{figure}[!t]
\centering
\includegraphics[width=1.0\linewidth]{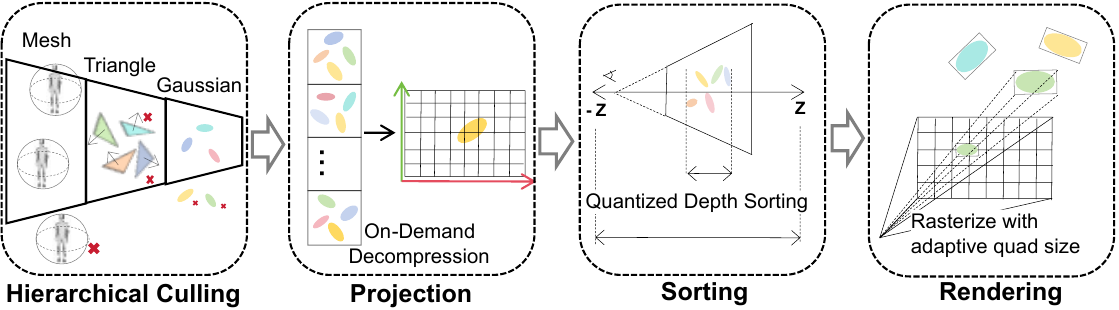}
\caption{GPU-Driven Rendering Pipeline for HRM$^2$Avatar.} \label{fig:rendering_pipeline}
\Description{Please see the caption.}
\end{figure}

To render high-fidelity Gaussian avatars on mobile devices, 
we address significant computational challenges posed by limited GPU capabilities and memory bandwidth. %
We propose a highly optimized GPU-driven rendering pipeline, as shown in Fig.~\ref{fig:rendering_pipeline}, specifically designed for mobile platforms, integrating data rearrangement, mesh-to-Gaussian hierarchical culling and single-pass stereo rendering particularly for AR/VR devices.

\paragraph{Data Rearrangement.} Data rearrangement includes two folds: compression and decompression of Gaussian attributes for reducing memory bandwidth, and depth quantization for fast Gaussian sorting.
Due to memory bandwidth limitation of mobile GPUs, direct memory access to uncompressed Gaussian data severely limits real-time rendering performance. Thus we introduce an offline compression and on-demand GPU-side data decompression scheme. 
Specifically, Gaussian attributes are compressed using the chunk-based compression~\cite{UnityGS2023Aras} after avatar reconstruction.
During runtime, a two-phase on-demand decompression is executed: 
positional data of all splats are decompressed initially for early visibility culling and full attribute decompression is performed exclusively for splats passing visibility tests. 
Traditional floating-point Gaussian depth sorting is computationally expensive on mobile GPUs. We introduce a quantized depth sorting scheme that maps view-space z-coordinates to a reduced precision integer range, enabling faster GPU sort operations~\cite{AMD2020FidelityFX} without perceptible visual degradation.

\paragraph{Mesh-to-Gaussian Hierarchical Culling.} 
Rendering invisible or negligible opacity Gaussians wastes GPU resources and negatively impact frame rates.
We develop a hierarchical visibility culling framework that exploits our mesh-Gaussian hybrid representation.
This three-tier culling system operates at mesh, triangle and splat levels to progressively reject invisible primitives, significantly reducing visible splat counts.
Specifically, mesh-level frustum culling uses bounding spheres to reject components outside the viewing frustum. Surviving meshes undergo triangle-level back-face culling based on view direction and triangle normals, leveraging the single-face visible Gaussian as detailed in supplementary material. 
At the splat level, visibility queries against associated triangles and per-splat frustum tests further reduce candidates. 

\paragraph{Single-Pass Stereo Gaussian Rendering.} Stereo rendering is critical for immersive VR experiences, but naively rendering both eyes independently nearly doubles the computational cost. 
To mitigate this, we implement single-pass stereo Gaussian rendering, 
where shared computations (e.g., skinning, data decompression) are executed once per frame and reused across eyes.
For view-dependent operations, culling is performed per-eye but share a unified visibility buffer to avoid redundant updates for common splats. Moreover, sorting is executed only using the left-eye camera and the result is shared to the right-eye because the forward directions of two eyes are nearly parallel on current AR/VR devices.
This approach reduces memory bandwidth usage and maintains real-time performance without perceptible quality loss.

\section{EXPERIMENT}
\subsection{Experimental Settings}
\paragraph{Implementation Details.} We use a single NVIDIA RTX 4090 GPU for training, with the optimization process comprising a total of 200k steps. Training takes about 7 hours for each subject. We set the hyper-parameters $\lambda_{L1}=0.8$, $\lambda_\text{ssim}=0.2$, $\lambda_\text{lpips}=0.1$, $\lambda_\text{mask}=1$, $\lambda_\text{sem}=1$, $\lambda_\text{collision}=5\times 10^{-4}$. All weights remain constant during training except the mask and LPIPS losses. During the first quarter of the steps, we up‑weight the mask loss to expedite silhouette alignment. The LPIPS loss is introduced at the 150k step to enhance fidelity. We also employ a progressive resolution strategy: images are rendered at 0.1$\times$ resolution for the first 100k steps, and gradually increased to full resolution from 100k to 175k steps. 

\paragraph{Datasets.} We evaluated HRM$^2$Avatar using five subjects captured with our protocol and four subjects (\textit{bike}, \textit{citron}, \textit{jogging}, and \textit{seattle}) from NeuMan~\cite{jiang2022neuman}. 
With our protocol, each subject was captured using an iPhone, result in 300-400 frames per subject with $1512 \times 2016$ resolution. 
The self-captured subjects exhibit diverse clothing types, including short- and long-sleeved tops, shorts, pants and skirts. 

\subsection{Comparison}
Tab.~\ref{tlb: comparisons on 2KMonoBody} and Fig.~\ref{custom} present a comparative evaluation of HRM$^2$Avatar against two state-of-the-art baselines, GaussianAvatar~\cite{hu2024gaussianavatar} and ExAvatar~\cite{moon2024exavatar}, on the self-captured datasets. 
We have partitioned 10\% of the full-body images to the test set, with the remaining for training.
The values presented in Tab.~\ref{tlb: comparisons on 2KMonoBody} are evaluated based on the self-driving images and their corresponding ground-truth images with foreground masks in the testing set. {\color{blue} Following the evaluation protocol of ExAvatar~\cite{moon2024exavatar}, we fit SMPL-X poses on the testing set to compute metrics, ensuring alignment of the major body parts.}
As shown in Fig.~\ref{custom}, while baseline methods exhibit plausible geometric structures at macro level (e.g., basic facial and body  topology), they fail to produce high-fidelity texture such as the logo and skin texture, as well as correct deformation for loose clothing, especially the skirts.
Our method achieves higher image quality on detailed texture and clothing dynamic, achieving realistic details at high resolution. 

\begin{table}[!h]
\renewcommand{\arraystretch}{1.1}
\centering
\caption{Comparisons on our dataset. Our method exhibits an unprecedented performance supremacy.}
\label{tlb: comparisons on 2KMonoBody}
\resizebox{0.9\columnwidth}{!}{ 
\begin{tabular}{l|ccc}
\toprule
Methods & \textbf{PSNR} $\uparrow$ & \textbf{SSIM} $\uparrow$ & \textbf{LPIPS} $\downarrow$ \\ \midrule
GaussianAvatar~\cite{hu2024gaussianavatar} &19.78&0.931&0.075 \\ 
ExAvatar~\cite{moon2024exavatar} &24.43&0.948&0.051 \\ %
Ours&\textbf{26.70}&\textbf{0.963}&\textbf{0.040} \\ \bottomrule
\end{tabular}
}
\end{table}

\begin{figure}[!t]
\centering
\includegraphics[width=1.0\linewidth]{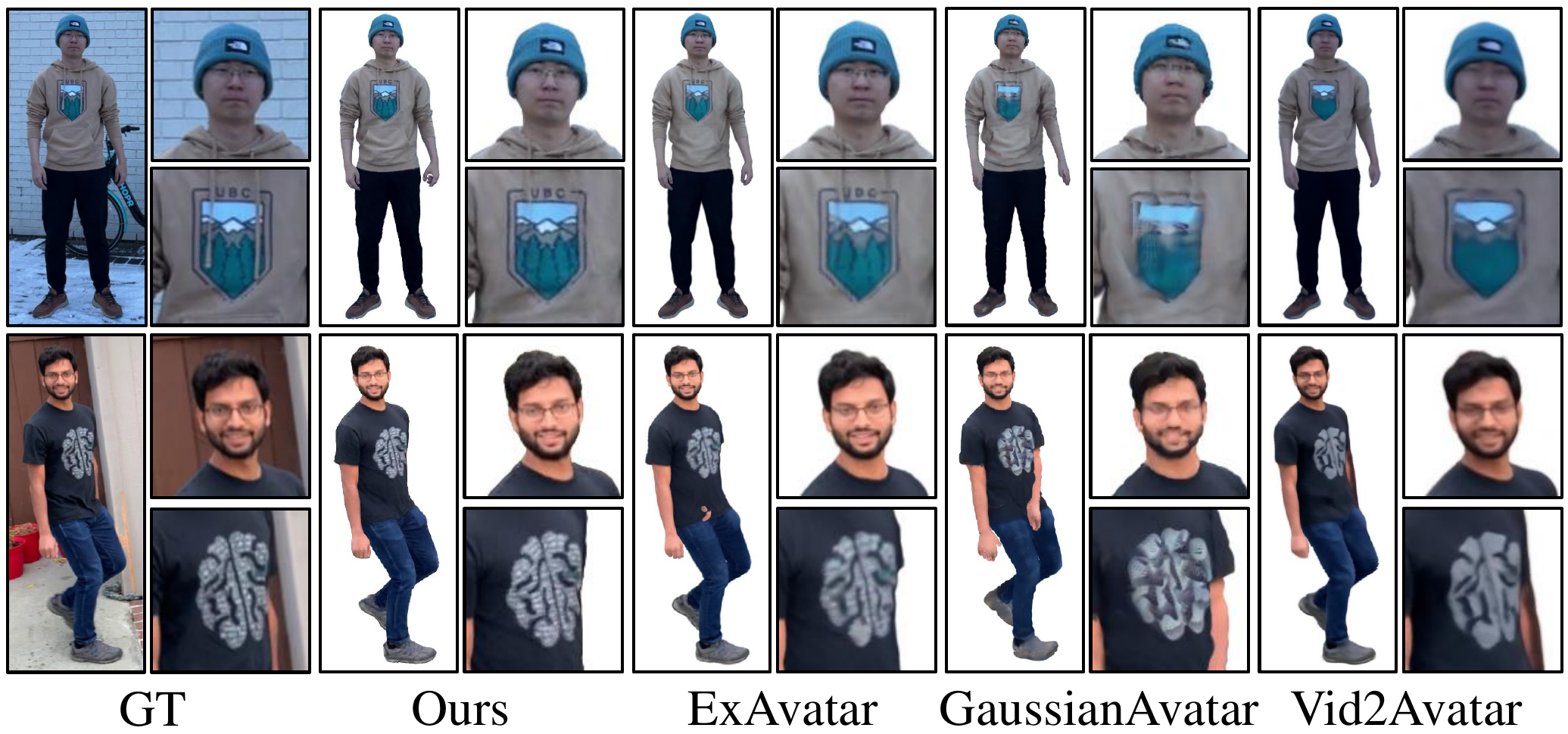}
\caption{Qualitative comparisons on NeuMan testset.}\label{neumantest}

\Description{Please see the caption.}
\end{figure}

In Tab.~\ref{tlb: comparisons on neuman} and Fig.~\ref{neumantest}, we compare our method with SOTA baselines on NeuMan dataset. The statistics are from original papers of Vid2Avatar-Pro~\cite{guo2025vid2avatarpro} and ExAvatar~\cite{moon2024exavatar}.
Following \cite{moon2024exavatar, Qian20243DGS-Avatar}, we fit SMPL-X parameters of testing frames while freezing all other parameters to evaluate quantitative metrics. Given that the Neuman dataset lacks static data and exhibits minimal relative motion between clothing and body, we streamline our approach to a single-layer representation by excluding clothing extraction, collision/semantic loss, and the static-dynamic co-optimization. By virtue of the mesh-driven hybrid representation and dynamic training strategy, HRM$^2$Avatar achieves the best metrics among these SOTA methods, and produces richer high-frequency details and more realistic wrinkle shadows.

\begin{table}[!h]
\renewcommand{\arraystretch}{1.1}
\centering
\caption{Comparisons on the NeuMan dataset. HRM$^2$Avatar outperforms all baseline methods.}
\label{tlb: comparisons on neuman}
\resizebox{\columnwidth}{!}{ 
\begin{tabular}{l|ccc}
\toprule
Methods & \textbf{PSNR} $\uparrow$ & \textbf{SSIM} $\uparrow$ & \textbf{LPIPS} $\downarrow$ \\ \midrule
NeuMan~\cite{jiang2022neuman} &29.32&0.972&0.014 \\
Vid2Avatar~\cite{guo2023vid2avatar} &30.70&0.980&0.014 \\
GaussianAvatar~\cite{hu2024gaussianavatar} &29.94&0.980&0.012 \\ 
3DGS-Avatar~\cite{Qian20243DGS-Avatar} &28.99&0.974&0.016 \\
ExAvatar~\cite{moon2024exavatar} &34.80&0.984&\textbf{0.009} \\ 
Vid2Avatar-Pro~\cite{guo2025vid2avatarpro} &32.71&0.983&0.012 \\ \midrule
ours w/o close-up, w/o clothing &\textbf{35.48}&\textbf{0.986}&0.011 \\ \bottomrule
\end{tabular}
}
\end{table}

\begin{figure}[!t]
  \centering
  \includegraphics[width=\columnwidth]{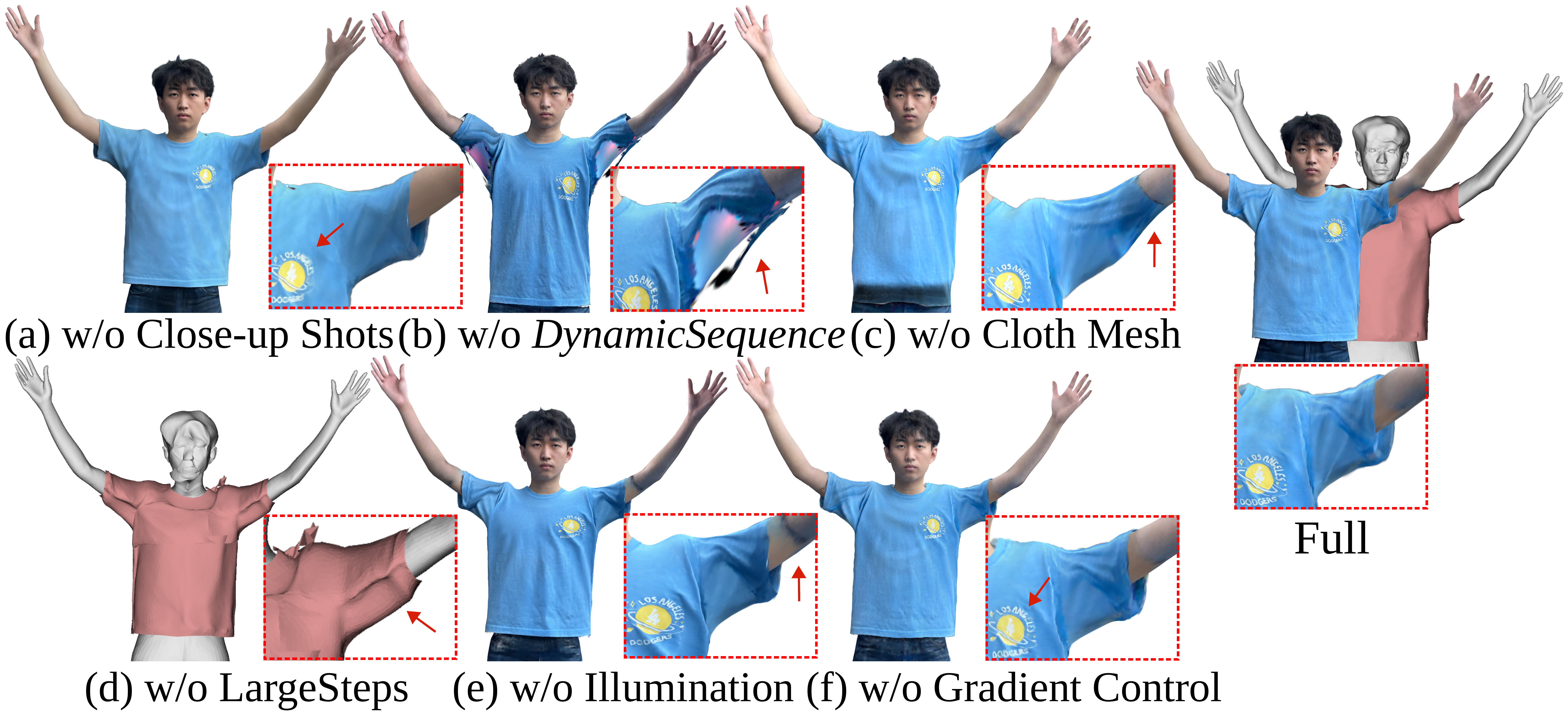}

  \resizebox{1.0\columnwidth}{!}{
    \begin{tabular}{lccc|lccc}
\toprule
 Version & \textbf{PSNR} $\uparrow$ & \textbf{SSIM} $\uparrow$ & \textbf{LPIPS} $\downarrow$& Version & \textbf{PSNR} $\uparrow$ & \textbf{SSIM} $\uparrow$ & \textbf{LPIPS} $\downarrow$ \\
\midrule
(a) w/o C.S. & 26.56 & 0.950 & 0.042 & (d) w/o L.S. & 26.07  & 0.943 & 0.050 \\
(b) w/o D.S. & 24.89  & 0.942 & 0.046 & (e) w/o Ill. & 26.14 & 0.950 & 0.043 \\
(c) w/o C.M. & 26.72 & 0.950 & 0.045 & (f) w/o G.C. & 26.72 & 0.952 & 0.043  \\
\textbf{Full} & \textbf{26.80}  & \textbf{0.952} & \textbf{0.041}  \\
\bottomrule
\end{tabular}
  }
  \caption{\textbf{Ablation Studies.} All metrics are measured on the test set of the male presented in the figure. %
  }\label{Ablation Study}
  \Description{Please see the caption.}
\end{figure}

\subsection{Ablation Studies}
We conduct ablation studies on the major factors that affect the final results as shown in Fig.~\ref{Ablation Study}, detailed as follows. 

\paragraph{Data-Related Ablation.} 
Our method integrates \textit{StaticSequence} (especially close-up shots) and \textit{DynamicSequence} for joint training. 
Fig.~\ref{Ablation Study}(a) demonstrates that the removal of close-up shots results in substantial degradation of fine-grained texture reconstruction, particularly the clothing logo, exhibiting unreality at high resolutions. Fig.~\ref{Ablation Study}(b) reveals that the ablation of \textit{DynamicSequence} results in significant deformation artifacts under novel driving poses, manifesting as sleeve penetration and loss of dynamic details in clothing lower edge movements.

\paragraph{Clothing-Related Ablation.} Our method constructs a \blue{clothed mesh-driven Gaussian avatar} representation in the geometry space to represent clothing dynamics, which is crucial for the motion realism at high resolutions. 
Fig.~\ref{Ablation Study}(c) illustrates that removing the \blue{cloth mesh} results in unnatural adhesion of clothing to the body surface and visible artifacts at the garment’s lower boundaries during arm elevation. This occurs due to the single-layer representation’s overfitting to rigid deformation patterns observed in the training data.
To mitigate geometric distortion during high-resolution training, we employ LargeSteps~\cite{nicolet2021large} to regularize clothing deformations. As demonstrated in Fig.~\ref{Ablation Study}(d), omitting LargeSteps results in geometric distortion due to insufficient constraints from monocular input.

\paragraph{Fidelity-Related Ablation.} Fig.~\ref{Ablation Study}(e) demonstrates that the absence of pose-conditioned illumination modeling induces abnormally black regions on the arm, and the wrong wrinkled shadows of the clothing. The reason is that while learning visual details from static sequences, illuminations are also learned into the SH coefficients. This phenomenon occurs because illuminations in \textit{StaticSequence} are inadvertently incorporated during the training progress. Our proposed gradient control strategy further improve the visual quality. As illustrated in Fig.~\ref{Ablation Study}(f), the exclusion of the gradient control reduces logo clarity relative to the complete model, yet still achieves superior definition compared to the close-up ablation (Fig.~\ref{Ablation Study}(a)). This progressive enhancement demonstrates the incremental efficacy of our hybrid capture and training methodology.

The quantitative metrics of the ablation studies are also summarized in the table in Fig.~\ref{Ablation Study}. 
For more ablation of minor factors such as non-rigid deformation MLP and losses, please refer to the supplementary materials.

\subsection{Runtime Performance}
We evaluated the runtime performance of the reconstructed avatar consisting of 533,695 splats on the iPhone 15 Pro Max and the Apple Vision Pro.
On the iPhone, we conducted tests at 2048$\times$945 resolution, with the avatar occupying the full screen. 
For the Vision Pro, we used its native 1920$\times$1824 per-eye resolution, positioning the avatar 2 meters from the user.

\begin{figure}[!t]
\centering
\includegraphics[width=0.9\linewidth]{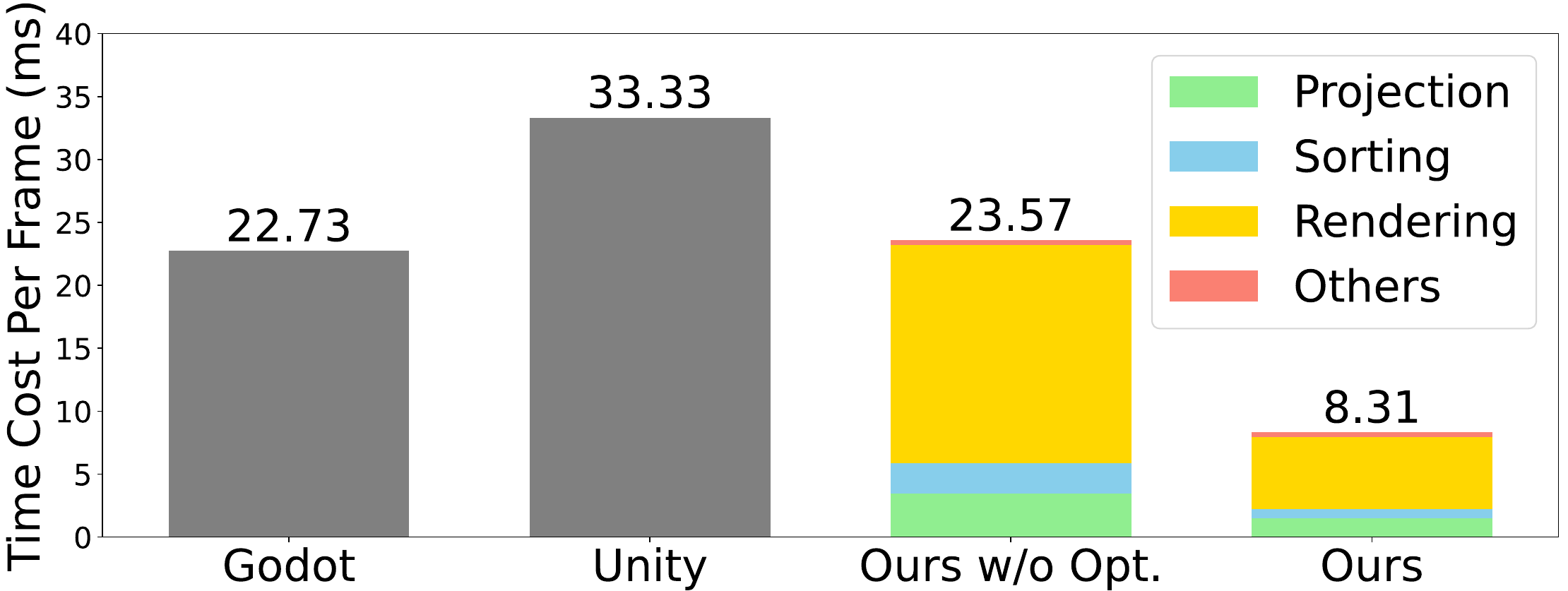}
\caption{\textbf{Runtime Performance on iPhone 15 Pro Max.}} \label{fig:runtime_performance}
\Description{Please see the caption.}
\end{figure}

We compared our optimized pipeline's runtime performance against the baselines (3DGS implementations in Godot~\cite{godotGS2023} and Unity~\cite{UnityGS2023Aras}) and our pipeline without optimizations (Ours w/o Opt.).
Results for the iPhone are presented in Fig.~\ref{fig:runtime_performance}, showing per-frame times for Godot/Unity, along with the time cost of each rendering pass. 
Note that the avatars in Godot/Unity are static, while ours supports dynamic user interaction. Comparisons on Apple Vision Pro were omitted because the Godot and Unity Gaussian Splatting implementations are not currently deployable to the device.
Overall, the optimized rendering pipeline achieves 120 FPS on the iPhone and 90 FPS on Apple Vision Pro, compared to 40 FPS and 39 FPS, respectively, without optimizations. 

\begin{table}[!h]
\renewcommand{\arraystretch}{1.1}
\centering
\caption{Ablation studies on runtime performance on iPhone 15 Pro Max and Apple Vision Pro. Time costs (ms) with optimization strategy off and on, and the speedup times.}
\label{tlb: ablation for runtime}
\resizebox{\columnwidth}{!}{ 
\begin{tabular}{l|ccc|ccc}
\toprule
 &  \multicolumn{3}{|c|}{iPhone 15 Pro Max} &\multicolumn{3}{|c}{Apple Vision Pro} \\
Strategy & OFF & ON & Speedup & OFF & ON & Speedup
\\ \midrule
Hierarchical Culling &15.24&8.31&1.83$\times$&15.79&10.38&1.52$\times$     \\ 
 On-demand Decompression& 2.87&1.48&1.93$\times$    & 2.60&1.98&1.31$\times$   \\
Depth Quantization &1.43&0.72&1.99$\times$          &1.06&0.56&1.88$\times$         \\ 
Single-Pass Stereo Rendering &  & N/A &             & 13.11 & 10.49 & 1.25$\times$ \\
\bottomrule
\end{tabular}
}
\end{table}

We also conduct ablation studies on the optimization strategies. The results are presented in Tab.~\ref{tlb: ablation for runtime}, evaluating the performance of these optimizations across task-specific metrics. 
For on-demand decompression, efficiency is measured by the execution time of the projection pass, while depth quantization performance is assessed by the sorting pass execution time. The total frame time cost of hierarchical culling and single-pass stereo rendering are also reported. Through chunk-based compression, we reduced the runtime memory footprint to 10\% of its original size while maintaining rendering quality. For more details on runtime memory, please refer to the supplementary material.

\section{LIMITATION AND FUTURE WORK}
Although HRM$^2$Avatar outperforms existing monocular full-body avatar methods, it still has several limitations. (1) Limited facial expressiveness. The current pipeline does not synthesize realistic expressions such as talking or laughing, because the scan protocol omits fine facial dynamics. Incorporating an additional monocular facial-expression sequence and fine-tuning a face-specific head network could address this gap. (2) Lack of dynamic hair modeling. Hairstyles that undergo significant non-rigid motion (e.g., swaying tresses) are currently treated as static Gaussians attached to the head mesh, so high-frequency hair dynamics are not captured at all. Future work may decouple a dedicated hair mesh from the body and design a more flexible Gaussian-binding strategy to simulate complex motion. {\color{blue} (3) The reconstructions may present some artifacts, particularly under large articulations, including unnatural clothing deformations, baked-in shadows, and occasional cloth-body interpenetration.
Capturing additional motion sequences with more diverse poses could help alleviate these artifacts, although achieving physically accurate garment dynamics and shadow disentanglement remains challenging with monocular input. A failure case of cloth-body interpenetration under large articulation is illustrated in the supplementary material.
(4) The current training pipeline is time-consuming, which may limit the efficiency of avatar creation. The training process for each subject takes 7 hours on a single GPU, with geometry optimization across 200-300 training poses being the primary computational bottleneck. Future improvements may include engineering optimizations and geometric priors from pre-trained models to accelerate reconstruction.}

\section{CONCLUSION}
We present HRM²Avatar, the first system that turns a single-phone scans into a high-fidelity, fully animatable avatar and achieves real-time interactive experiences on mobile devices.
Its key ingredients are two-stage capture which contains \textit{StaticSequence} for detail textures and \textit{DynamicSequence} for motion from an ordinary phone.
We adopt clothed mesh-driven Gaussian avatar representation, and equip it with pose-dependent geometrical deformation and illumination variation to model the animation and shading of avatar. 
GPU-driven Gaussian rendering pipeline with data rearrangement, hierarchical culling and single-pass stereo rendering is developed to guarantee high-res and high-performance rendering on mobile devices.
Experiments show our method achieve better visual quality, motion accuracy, and frame rate than prior monocular methods. 

\begin{acks}
We thank the anonymous reviewers for their valuable feedback and constructive suggestions used to improve this paper. 
We would also like to express our sincere gratitude to Yan Zhang, Luwei Xu, Ningwei Zhou, Junjian Li, Yuan Peng, Qiang Zhou, Qian Shen, Pengwei Guo, and Lijie Zhang for their contributions to the system realization, experimental data collection, and demo implementation.

\end{acks}

\clearpage
\balance
\bibliographystyle{ACM-Reference-Format}
\bibliography{reference}

\clearpage
\appendix
\onecolumn

\section*{}
\begin{figure*}[!t]
\centering
\includegraphics[width=1.0\linewidth]{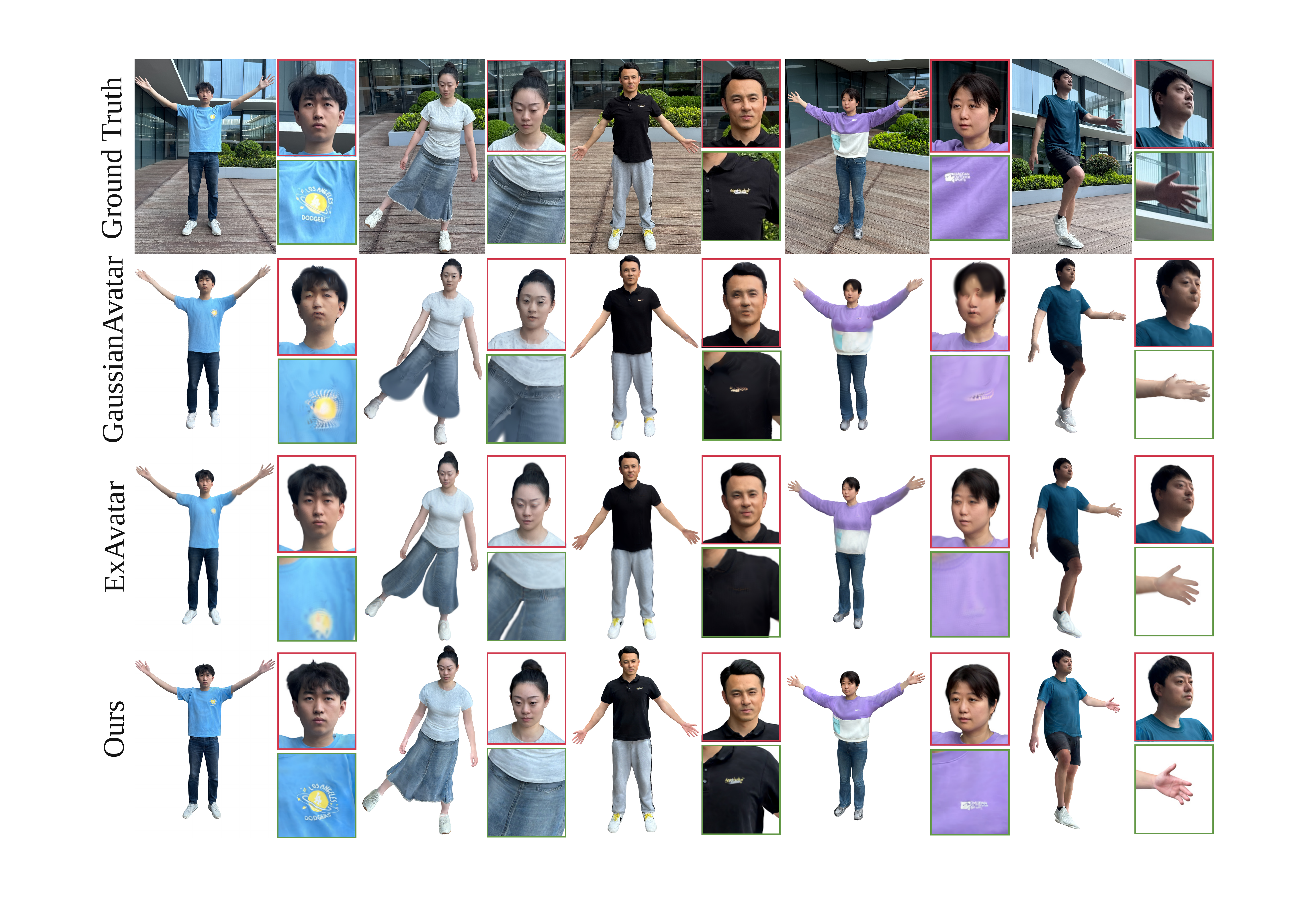}
\caption{Self-driven animation comparzisons with monocular avatar methods.}\label{custom}

\Description{Please see the caption.}
\end{figure*}

\begin{figure*}[!t]
\centering
\includegraphics[width=0.85\linewidth]{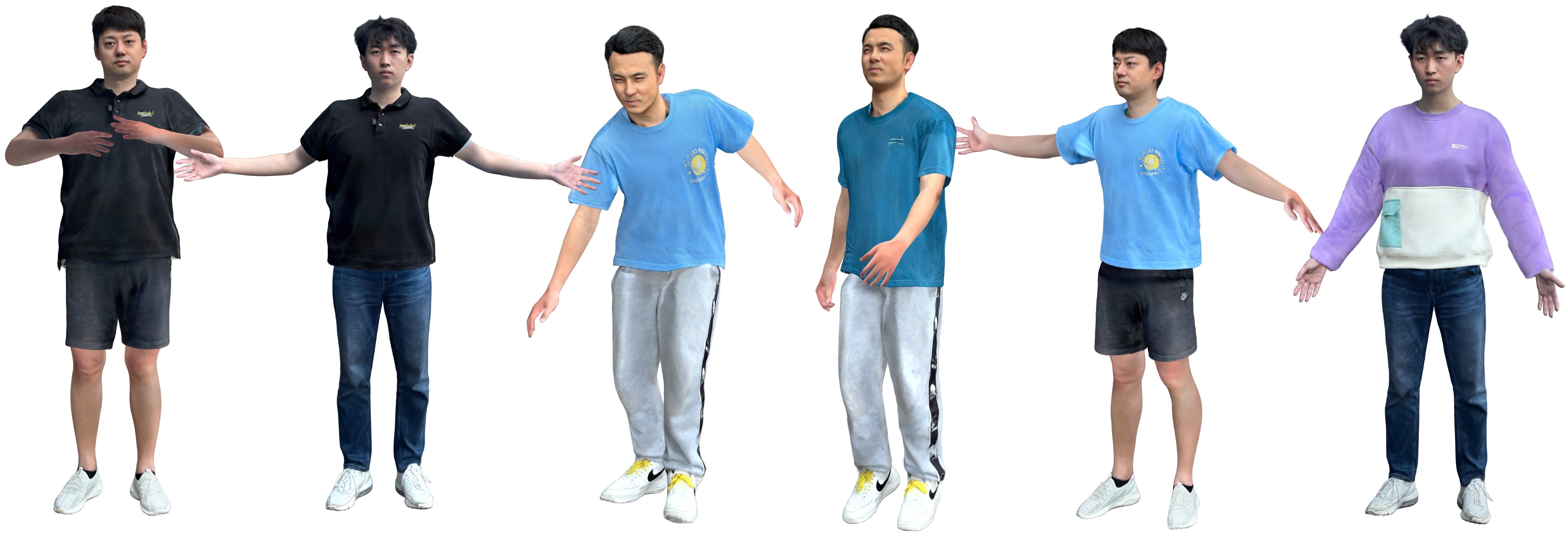}
\caption{\textbf{An example of cloth exchange.} We achieved realistic garment transfer from one subject to another through a simple collision-aware positional refinement, demonstrating promising opportunities for virtual try-on applications.}\label{clothesExchange}
\Description{Please see the caption.}
\end{figure*}

\begin{figure*}[!t]
\centering
\includegraphics[width=0.95\linewidth]{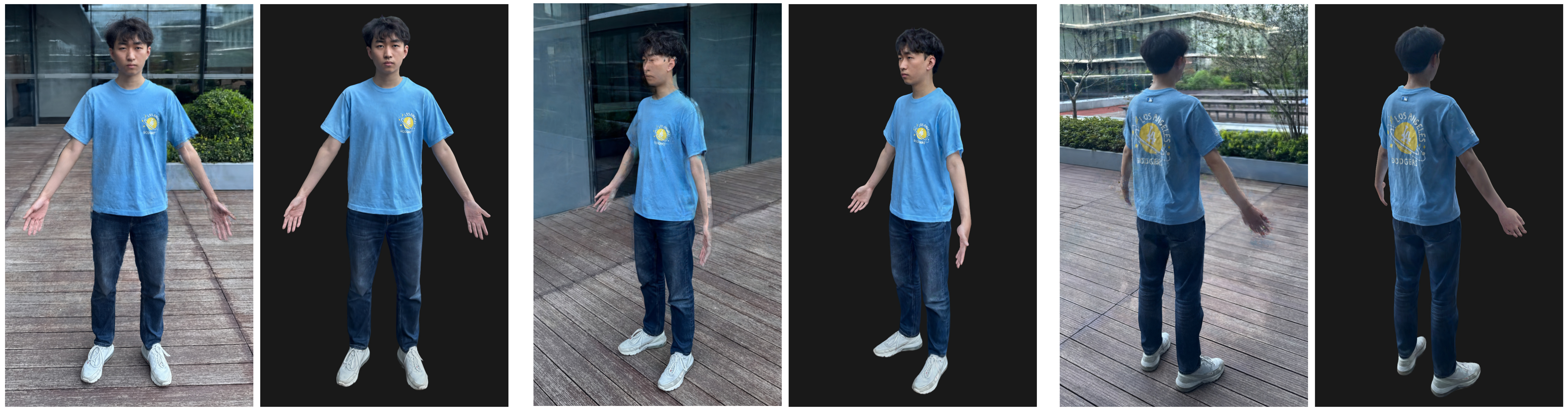}
\caption{\textbf{Deviation in \textit{StaticSequence}.} Due to the slight movement of the human body during monocular capturing, \textit{StaticSequence} cannot be simply treated as a multi-view scene. Images with outdoor-background are reconstructed via native 3DGS on \textit{StaticSequences}, whereas black-background images is the results of our method. The 3DGS's results exhibit artifacts on hands and clothing logos, which are induced by subtle motions. Our optimization strategy solve this issue.}\label{DeviationInStaticSequence}

\Description{Please see the caption.}
\end{figure*}

\end{document}